\documentstyle[aps,epsf]{revtex}
\input amssym.def

\begin{document}
% \draft command makes pacs numbers print
%\draft
\author{
C. Chicone\thanks{ Department of  Mathematics,
University of Missouri, Columbia, MO 65211.
Supported by the NSF grant  DMS-9531811.} ,\quad
B. Mashhoon\thanks{ Department of  Physics and Astronomy,
University of Missouri, Columbia, MO 65211.
}\quad
and D. G. Retzloff\thanks{ Department of Chemical Engineering,
University of Missouri, Columbia, MO 65211.}
}
\date{\today}
\title{Gravitational Ionization: A Chaotic Net in the Kepler System}
\maketitle
\begin{abstract}   The long term nonlinear dynamics of a Keplerian binary
system under the combined influences of gravitational radiation damping and
external tidal perturbations is analyzed.  Gravitational radiation reaction
leads the binary system towards eventual collapse, while the external periodic
perturbations could lead to the ionization of the system via Arnold diffusion.
When these two opposing tendencies nearly balance each other, interesting
chaotic behavior occurs that is briefly studied in this paper.  It is possible
to show that periodic orbits can exist in this system for sufficiently small
damping.  Moreover, we employ the method of averaging to investigate the
phenomenon of capture into resonance.
\end{abstract}
%\pacs{}
\section{Introduction}
The Kepler problem traditionally involves two point masses $m_1$ and
$m_2$ moving under the influence of their mutual gravitational interaction.  The
incorporation of relativistic gravitational effects in the Kepler system brings
about three main post-Newtonian modifications in the traditional picture:

(i) There exist gravitoelectromagnetic post-Newtonian effects that need to be
taken into account.  The most important of these is the periastron
precession that has played an important role in the historical development of
general relativity.

(ii)  Moreover, the system emits gravitational radiation and hence there exist
{\it radiative} post-Newtonian effects in the orbit that are
characteristic of gravitational radiation damping.  This radiative damping is
consistent with the observed rate of inward spiraling in the Hulse-Taylor binary
pulsar system \cite{hulse,taylor}.

(iii)  The system is expected to be affected by gravitational waves that have
been emitted by other systems as well as by their post-Newtonian tidal perturbations.  
Gravitational radiation has not yet been
directly observed; however, the Hulse-Taylor binary pulsar data appear to be
consistent with the notion that gravitational radiation is emitted by binary
systems
\cite{taylor}.  Half of all stars are expected to be members of binary or
multiple systems that could emit gravitational waves; therefore, there might be
a cosmic background of gravitational radiation that has been generated by
various sources throughout the history of the universe.  Moreover, there could
also be primordial gravitational waves left over from the epoch at which the
Hubble expansion began.
The inclusion of all of these effects in the Kepler problem is impractical; in
fact, the two-body problem in general relativity is intractable.  To render the
problem amenable to mathematical analysis, it is therefore necessary to replace
the actual problem by a model that contains the main physical effects of
interest.  We consider a Kepler system in which the main effects of emission and
absorption of gravitational radiation are taken into consideration.  Gravitation
is a spin-2 field; therefore, the main radiative effects first occur in general
at the quadrupole level in emission as well as in absorption.  At this level,
the wavelength of the radiation is large compared to the size of the system.
In this work, we analyze the radiative perturbations of a Keplerian binary
system in the quadrupole approximation.

In our previous papers \cite{cmr,cmr1}, we considered the problem of
ionization  of a Keplerian binary system by a normally incident periodic
gravitational wave.  To render the absorption problem tractable, we simply
ignored the emission of gravitational radiation by the binary system.
Therefore, the influence of the gravitational radiation reaction on the binary
orbit was not taken into account.  In the absence of this dissipative effect,
our model of gravitational ionization turned out to be a Hamiltonian system to
which the basic results of the Kolmogorov-Arnold-Moser~(KAM) theory could be
applied under certain circumstances.  The main purpose of the present work is
to take gravitational radiation damping into account.  Thus, we study in this
paper the long term nonlinear evolution of a Keplerian binary system under the
combined action of perturbations due to the emission and absorption of
gravitational radiation.

The nonlinear dynamics of a Keplerian binary that emits and absorbs
gravitational radiation is given in our model by
\begin{equation}\label{BasicEq}
\frac{d^2r^i}{dt^2} + \frac{kr^i}{r^3} +  {\cal R}^i = -\epsilon {\cal
K}_{ij}(t)\: r^j,
\end{equation}
where ${\bf r}:= {\bf x}_1-{\bf x}_2$ denotes the relative
two-body orbit, $r$ is the  length of $\bf r$, and $k = G_0(m_1+m_2)$.  Here the
radiation reaction term
${\cal R}$ is given by
\begin{equation}\label{RadiationReaction}
{\cal R} = 
\frac{4G_0^2m_1m_2}{5c^5r^3}\left[\left(12 v^2-30\dot{r}^2-
                                                    \frac{4k}{r}\right){\bf v} 
  -\frac{\dot{r}}{r}\left(36v^2-50\dot{r}^2+\frac{4k}{3r}\right){\bf r}\right],
\end{equation}
where ${\bf v}$ is the relative velocity, $G_0$ is Newton's constant, $c$ is
the speed of light in vacuum, and an  overdot represents differentiation with
respect to time, i.e. $\dot{r} = dr/dt$.  We emphasize here that
equation (\ref{BasicEq}) represents the simplest equation for the relative
motion of two point masses $m_1$ and
$m_2$ that contains the main dynamical effects of emission and absorption of
gravitational radiation that we wish to study here; in fact, other
post-Newtonian contributions have been simply ignored.  
The justification for
this approach emerges from our analysis; that is, the substitution of the
actual intractable problem by our simple model permits us to arrive at 
interesting results of possible astrophysical interest.

Let the background spacetime metric be given by $g_{\mu\nu} = \eta_{\mu\nu} +
\epsilon \chi_{\mu\nu}$, where $\eta_{\mu\nu}$ denotes the Minkowski metric,
$\epsilon$ is a small parameter indicative of the strength of the external
radiation field $(0 < \epsilon << 1)$, and $\chi_{\mu\nu}$ represents the
gravitational radiation field pervading the space.  We impose the
transverse-traceless gauge condition such that 
$\chi_{0\mu}=0$, ${\partial}_j \chi_{ij} = 0$, and
$(\chi_{ij})$ is traceless; moreover, $\chi_{ij}$ satisfies the wave equation
$\Box \chi_{ij} = 0$.  Imagine now that the mutual gravitational interaction
between the two masses is turned off and they are therefore following geodesics
of the background spacetime manifold.  The spatial separation between the two
masses is assumed to be small compared to the wavelengths of the background
(external) waves; therefore, the relative motion of the two masses can be
expressed via the Jacobi equation.  It is useful to express the deviation
equation with respect to a Fermi coordinate system established along the
worldline of the center of mass of the system.  The Fermi system is the most
natural extension of a local inertial frame along the path of an observer in
spacetime.  Hence, the relative acceleration of the two bodies in the Fermi
system takes on a Newtonian form that is derivable from a quadrupole potential
given by $\frac{1}{2} \epsilon {\cal K}_{ij}(t) x^ix^j$ with
\begin{equation}\label{QuadPot}
{\cal K}_{ij}(t) = -\frac{1}{2}\frac{\partial^2\chi_{ij}}{\partial t^2} (t,{\bf
x}_{\mbox{\rm \tiny CM}})\;,
\end{equation}
where $(m_1+m_2){\bf x}_{\mbox{\rm \tiny CM}} = m_1{\bf x}_1 + m_2{\bf x}_2$.

In equation  (\ref{BasicEq}), we have neglected terms of  order $\epsilon^2$,
$\epsilon\delta$, $\delta^2$, or higher, where $\delta$ is the strength of the
radiation reaction term as defined in Appendix~\ref{DEQM}; 
that is, in conformity with the Newtonian appearance of
equation (\ref{BasicEq}) the various forces have been linearly superposed.  A
detailed treatment of equation (\ref{BasicEq}) with
${\cal R} = 0$ is contained in our previous papers \cite{cmr,cmr1}, which
considered only normally incident external waves for the sake of simplicity.
We shall also choose here an external periodic monochromatic gravitational wave
that is perpendicularly incident on the orbital plane.  Let this plane be
characterized by the vanishing of the third component of ${\bf r}$. 
Then, the nonzero elements of the tidal matrix 
$(\epsilon {\cal K}_{ij})$ are given by
\begin{eqnarray}\label{TidalMatrix}
{\cal K}_{11} & = & - {\cal K}_{22} = \alpha\Omega^2\cos(\Omega t), \nonumber \\
{\cal K}_{12} & = &  {\cal K}_{21} = \beta\Omega^2\cos(\Omega t + \rho),
\end{eqnarray}
where $\alpha$ and $\beta$ are of the order of unity and represent the constant
amplitudes of the two independent linearly polarized components of the incident
monochromatic wave of frequency $\Omega$ and $\rho$ is the constant phase
difference between the two components (cf. \cite{cmr}).

It is interesting to note 
the partial reciprocity between the emission and absorption of gravitational
radiation.  In the quadrupole approximation, an elliptical
Keplerian binary system emits gravitational waves only at frequencies $m\omega$,
$m= 1, 2, 3, \cdots$, where $\omega$ is the Keplerian frequency of the orbit
\cite{peters}.  It follows from a linear perturbation analysis \cite{mashoon1}
that when a Newtonian binary system absorbs gravitational energy from an
incident wave of frequency $\Omega$, resonances occur at $\Omega = m\omega$,
$m= 1, 2, \cdots$.  However, it is important to point out that while the binary
monotonically loses energy in the form of gravitational radiation, the
absorption of gravitational radiation energy is not monotonic.  {\it The
system can gain or lose energy in absorption}.  In our previous papers
\cite{cmr,cmr1}, we examined the long term nonlinear evolution of the dynamical
system with ${\cal R} = 0$ and proved the possibility of existence of periodic
orbits for which the net flow of energy between the binary and the external
periodic gravitational radiation field must vanish.  In the present work, we
extend our previous results by taking due account of gravitational radiation
damping; in fact, periodic orbits are shown to exist for sufficiently small
$\delta/\epsilon$

The issue of {\it gravitational ionization} provided the original
motivation for our work.  The possibility of ionization of a Keplerian binary by
incident gravitational waves had been first discussed within the framework of
{\it linear} perturbation analysis \cite{mashoon1}.  However, the linear
perturbation treatment breaks down over time as a consequence of the appearance
of secular terms in the analysis.  On the other hand, the results of the
linear analysis can be applied to the possibility of  detection --- via the
solar system --- of gravitational pulses
\cite{mashoon2} or a stochastic cosmological background of gravitational waves
\cite{mashoon3}; a review of this general topic can be found in
\cite{grishchuk} and further work in this direction is contained in
\cite{Linet,Chau,Petukhov,Ivashchenko,Sbytov} and the references cited
therein.  Furthermore, the tidal influence of very long wavelength gravitational
waves on galaxies has been studied in \cite{Matsuda}.  Our {\it nonlinear}
perturbation treatment \cite{cmr,cmr1}  has led to the conjecture that
gravitational ionization is tantamount to Arnold diffusion.  Moreover, there
are interesting circumstances --- other than periodic orbits --- for which
ionization does not occur.  For instance, it follows from the KAM theorem that
when the incident wave is circularly polarized (i.e.
$\alpha=\beta$ and $\rho=\pm \pi/2$) the perturbed binary is forever confined
and ionization does not occur if $\epsilon$ is below a certain limit $(\epsilon
< \epsilon_{\mbox{\rm \tiny KAM}})$.  It is the purpose of this paper
to determine how
all of these results are affected by the presence of gravitational
radiation damping.  While our previous work \cite{cmr,cmr1} has involved
time-dependent (i.e. nonconservative) Hamiltonian systems with chaotic behavior,
the introduction of radiation damping leads to a dissipative dynamical system
that is not Hamiltonian.

At each instant of time $t$, the state of relative motion is determined by the
relative position and velocity $({\bf r}, {\bf v})$; however, it is useful to
employ instead the six orbital parameters of the osculating ellipse.  That is,
the unperturbed bounded motion is in general a Keplerian ellipse; therefore, it
is interesting to describe the state of relative motion at each instant by the
ellipse that the system would follow if the perturbations were turned off at
that instant.  The orbital elements of the osculating ellipse are closely
related to the Delaunay elements employed in this paper.   They are
particularly useful in a Hamiltonian system \cite{cmr,cmr1}, since the
transformation to the Delaunay action-angle variables 
is canonical and the corresponding
generating function is time-independent so that the magnitude of the
Hamiltonian is unchanged under the transformation.

A limiting case of the relative osculating ellipse is a circle.  In this case,
the system emits gravitational waves with a single frequency $2\omega$.  On the
other hand, linear perturbation theory reveals that resonant absorption occurs
at wave frequencies $\omega$, $2\omega$, and $3\omega$. However, it should be
remarked that after the passage of time the orbit will no longer be circular.
The nonlinear perturbation theory employed in our work is based on the ideas
and methods originally developed by Poincar\'{e} \cite{poincare}. 
An important aspect of this theory is its {\it lack of emphasis} on the precise
form of the {\it initial conditions}; clearly, this characteristic of the theory
is rather useful in celestial mechanics.
In this approach, the dynamical behavior of a physical 
system is investigated after
transients have died away and a steady state has been established.

In connection with the issue of initial conditions, it is necessary to remark
that the external periodic perturbation in our model is meant to represent the
dominant component of an initial wave packet composed mainly of long-wavelength 
Fourier
components consistent with the quadrupole approximation under consideration
here.  On the other hand, high-frequency waves with wavelengths much smaller
that the semimajor axis of the system are expected to have negligible influence
on the dynamics.  That is, along any significant portion of the relative orbit
the waves are expected to go through many oscillations and their net influence
is on average expected to be vanishingly small.  In our model, the strength of
the external perturbation is $\epsilon$; at present, it is expected that
$\epsilon \sim 10^{-20}$, though gravitational waves have not yet been detected
in the laboratory.  On the other hand, the strength of radiative damping in our
model is given by $\delta$ (cf. Appendix \ref{DEQM}); for instance, for the
binary
pulsar PSR 1913+16, discovered by Hulse and Taylor~\cite{hulse}, 
$\delta \sim 10^{-15}$, while for the Earth-Sun system $\delta \sim 10^{-26}$.  
The more
relativistic a binary system is, the larger the magnitude of $\delta$, which
for a physical system must always be less than $(20\sqrt{2})^{-1}$ as discussed
in Appendix \ref{DEQM}.

Throughout this paper, the external perturbing force is assumed to be due to
background gravitational waves.  However, as explained in detail in our first
paper \cite{cmr}, this is not necessary since any tidal influence on the binary
orbit by a far-away mass would contribute a forcing term with the same
functional form as in the equation of  relative motion
(\ref{BasicEq}).  In general, the binary system cannot be isolated from the
other masses in the universe; this is simply a result of the universal character
of gravitational attraction.  Thus equation (\ref{BasicEq}), which is
essentially the damped Newton-Jacobi equation, should contain the tidal
influence of distant masses as well as background gravitational waves in the
forcing function.  That is, the background spacetime curvature --- represented
in equation (\ref{BasicEq}) by the tidal matrix --- is due to the combined
gravitational influence of background waves as well as masses.
The analysis of the general problem  would involve the three-dimensional
Kepler system.  In our {\em planar} model,
we neglect the other masses and choose a normally incident monochromatic plane
wave for the sake of simplicity.

An interesting feature of equation (\ref{BasicEq}) must be noted here: 
The vector cross product of this equation with the relative position vector
results in
an equation for the rate of change of the orbital angular momentum with time.
In this result, the torque due to the gravitational radiation reaction is
proportional to the orbital angular momentum according to equation
(\ref{RadiationReaction}); therefore, the action of this torque is only to
change the angular momentum normal to the orbital plane. The torque due to the
external force is similar in character, since the external wave is transverse
and is assumed to be normally incident on the orbital plane.  The net result is
that the relative motion is {\it planar}, and this fact considerably simplifies
the analysis of the relative motion.  Moreover, in conformity with the
quadrupole approximation the net velocity of the center of mass (i.e. the
velocity of the binary system as a whole) is uniform.  Therefore, we choose
the ``inertial" reference frame $(t, {\bf x})$ for our model such that it is
comoving with the whole binary system and that ${\bf x}_{\mbox{\rm \tiny
CM}}=0$.  In this frame, the orientation of the coordinate system is so chosen
that the perturbed Keplerian motion has a positive initial orbital
angular momentum.

The motion in our model is planar; therefore, it is interesting to express
equation (\ref{BasicEq}) in terms of polar coordinates $(r, \theta)$ in the
orbital plane.  Thus,
\begin{eqnarray}\label{MathEQM}
\frac{dr}{dt} & = & p_r, \nonumber \\
\frac{d\theta}{dt} & = & \frac{p_{\theta}}{r^2}, \nonumber \\
\frac{dp_r}{dt} & = & -\frac{k}{r^2} + \frac{p_{\theta}^2}{r^3} +
\frac{16G_0^2m_1m_2}{5c^5}\frac{p_r}{r^3}\left(p_r^2+6\frac{p_{\theta}^2}{r^2}+
\frac{4}{3}\frac{k}{r}\right) \nonumber \\ & & \hspace*{.25in} -\epsilon r
\Omega^2 [\alpha\cos{2\theta}\cos{\Omega t}+\beta\sin{2\theta}\cos{(\Omega
t+\rho)}], \nonumber \\
\frac{dp_{\theta}}{dt} & = & \frac{8G_0^2m_1m_2}{5c^5}\frac{p_{\theta}}{r^3}
\left( 9p_r^2 - 6 \frac{p_{\theta}^2}{r^2} + 2\frac{k}{r}\right) \nonumber \\ &
& \hspace*{.25in} +\epsilon r^2 \Omega^2 [\alpha\sin{2\theta}\cos{\Omega
t}-\beta\cos{2\theta}\cos{(\Omega t+\rho)}].
\end{eqnarray}
Here $p_{\theta}$ is the orbital angular momentum and the initial conditions
are chosen such that $p_{\theta}>0$.  It is clear from these equations that the
corresponding vector field in  phase space is periodic with frequency
$\Omega$.

The main dynamical result of our work may now be stated  qualitatively
in terms of the characteristic forms of the perturbing functions.  The
gravitational radiation reaction force is proportional to $r^{-q}$ with $q
\geq 3$.  Thus if $r$ becomes small,  radiative damping takes over and
leads the system inexorably to collapse.  On the other hand, the external
potential is proportional to $r^2$ --- i.e. the energy exchange is proportional
to the orbital area  through which the normally incident wave passes (cf.
\cite{mashoon2}) --- so that if $r$ becomes large the external force takes over
and the results of our previous work \cite{cmr,cmr1} imply that Arnold
diffusion might take place leading to the ionization of the binary system.
Thus it appears that the long time dynamical behavior of the system under 
consideration here is characterized by two possibilities:
collapse to the origin and unbounded growth; in fact,
our results indicate that the collapse scenario is prevalent in most cases.

The plan of this paper is as follows:  In Section 2, we discuss 
the radiation reaction term ${\cal R}$ in equation (\ref{BasicEq}).  Section
3 is devoted to the determination of the bifurcation function and the proof of
the existence of periodic orbits in the damped system.  In Section 4, the
average behavior of the dynamical system (\ref{BasicEq}) with $\epsilon = 0$ is
determined for all times.  Section 5 describes the average properties of the
system (\ref{BasicEq}).  Section 6 considers the case of circularly polarized
incident waves.  Section 7 describes the chaotic net using numerical analysis.
A number of detailed computations are relegated to the Appendices.  The paper
relies on our previous work
\cite{cmr,cmr1} for background material regarding the topic of {\it
gravitational ionization}; however, we have tried to make this paper essentially
self-contained.
\section{Gravitational Radiation Damping}
In the quadrupole approximation under consideration here,  gravitational
waves carry energy and angular momentum away from the system but not linear
momentum.  Similarly, in absorption the orbit exchanges energy and angular
momentum with the incident gravitational wave but not linear momentum.  This
implies that the motion of the system as a whole --- i.e. the center-of-mass
motion --- is not affected by the emission and absorption of radiation in the
quadrupole approximation.  Taking proper account of the emission of
gravitational radiation,  the conservation laws of energy and momentum require
that these losses be reflected in the motion of the system.  To satisfy this
requirement, it is sufficient to include in the equations of motion of a
particle in the system  a gravitational radiation reaction force per unit
inertial mass given by
${\cal A}$,
\begin{equation}\label{RadReaForce1}
{\cal A}_i = -\frac{2 G_0}{15 c^5}\frac{d^5 {\cal D}_{ij}}{dt^5}x^j,
\end{equation}
where $({\cal D}_{ij})$ is the quadrupole moment of the system (cf. \cite{ww}
and the references cited therein).  We may look upon this force as the spin-2
analog of the standard spin-1 radiation reaction force in electrodynamics
\cite{lorentz}.  The radiation reaction term in the equations of motion cannot
be derived from a potential; otherwise, the dissipative dynamical system would
be Hamiltonian.

Let ${\cal E}$ denote
the energy radiated as gravitational waves; then,
\begin{equation}\label{EnergyEq}
\frac{d{\cal E}}{dt} = \frac{G_0}{45c^5}\stackrel{...}{{\cal D}}_{ij}
\stackrel{...}{{\cal D}}^{ij},
\end{equation}
where
\begin{displaymath}
{\cal D}_{ij} = \int \tilde\rho(3x^ix^j-\delta_{ij}x^2) dV.
\end{displaymath}
Here the quadrupole moment of the system is traceless by definition, $\tilde\rho$ is
the mass density, and $dV$ is the volume element.  Equation~(\ref{EnergyEq})
is a standard result of general relativity in the quadrupole approximation;
that is, it can be obtained from a detailed treatment of the
linearized form of the gravitational
field equations together with the physical interpretation of the
Landau-Lifshitz pseudotensor. Following this approach,
a similar expression can be derived in the quadrupole approximation
for the angular momentum carried away by the gravitational waves.
The energy and angular momentum radiated away via gravitational waves
are lost by the orbit;
therefore, the equation of orbital motion should reflect this loss.
For both energy and angular momentum,
this must be accomplished by the
introduction of the radiation reaction force. By analogy with
electrodynamics, we expect that the radiation reaction force  per unit inertial
mass is of the form given by equation (\ref{RadReaForce1}), so that for the
binary system under consideration  we can
express the equations of motion as
\begin{equation}\label{RadEq1}
\frac{d^2x^i_1}{dt^2} + \frac{G_0m_2({\bf x}_1-{\bf x}_2)^i}{r^3}  =
-\frac{2G_0}{15c^5}\frac{d^5{\cal D}_{ij}}{dt^5} x^j_1 -\epsilon {\cal
K}_{ij}(t) x_1^j,
\end{equation}
\begin{equation}\label{RadEq2}
\frac{d^2x^i_2}{dt^2} + \frac{G_0m_1({\bf x}_2-{\bf x}_1)^i}{r^3}  =
-\frac{2G_0}{15c^5}\frac{d^5{\cal D}_{ij}}{dt^5} x^j_2 -\epsilon {\cal
K}_{ij}(t) x_2^j,
\end{equation}
where ${\bf r} = {\bf x}_1-{\bf x}_2$. Multiplying the first
equation by $m_1$ and the second equation by $m_2$ and adding them results in
\[
\frac{d^2 x^i_{\mbox{\rm\tiny CM}}}{dt^2}=-\left[
\frac{2G_0}{15c^5}\frac{d^5{\cal D}_{ij}}{dt^5}
   +\epsilon {\cal K}_{ij}(t) \right] x^j_{\mbox{\rm \tiny CM}}.
\]
In the absence of perturbations, ${\bf x}_{\mbox{\rm \tiny CM}}$ is
at the origin of the coordinate system; therefore, this equation of 
motion for ${\bf x}_{\mbox{\rm \tiny CM}}$ 
implies that it is effectively unchanged
in our approximate treatment.
It is thus  possible to set 
${\bf x}_{\mbox{\rm \tiny CM}} = 0$, 
so that the emission or absorption of gravitational waves does
not affect the total linear momentum of the system in the quadrupole
approximation.    
The relative motion is
obtained by subtracting the two equations:
\begin{equation}\label{RadReaEq}
\frac{d^2 r^i}{dt^2} + \frac{G_0(m_1+m_2)r^i}{r^3} = 
  -\frac{2 G_0}{15 c^5}\frac{d^5{\cal D}_{ij}}{dt^5} r^j 
    -\epsilon {\cal K}_{ij}(t) r^j.
\end{equation}
It follows that the full content of the equations of motion is simply contained
in the expressions ${\bf x}_1 = m_2{\bf r}/M$ and ${\bf x}_2 = -m_1{\bf r}/M$,
where $M = m_1 + m_2$ is the total mass and ${\bf r}$ is given by the equation
of relative motion (\ref{RadReaEq}). The quadrupole moment of the system is then
given by
\begin{equation}\label{QuadMom}
{\cal D}_{ij} = m_1(3x^i_1x^j_1-\delta_{ij}x^2_1) + m_2(3x^i_2x^j_2-\delta_{ij}
x^2_2) = \mu(3r^ir^j-\delta_{ij}r^2),
\end{equation}
where $\mu$ is the reduced mass, i.e. $\mu = m_1m_2/M$.

It is clear from the form of equations (\ref{RadReaEq}) and (\ref{QuadMom}) that
the relative motion must be determined from an equation in which the order of
differentiation with respect to time exceeds two.  On the other hand, the higher
derivative terms only appear in the radiation reaction acceleration that has an
overall dimensionless strength given by a small parameter $\delta$ that is
particularly small for Keplerian (i.e. nonrelativistic) binaries as discussed in
Appendix \ref{DEQM}.  It follows that the problem can be avoided in this case if
we proceed iteratively and substitute for $\ddot{r}^i$ from equation~(\ref{RadReaEq})
every time it appears in the calculation of gravitational radiation damping.  
Then, keeping
only terms of the first order in the perturbation parameters $\epsilon$ and
$\delta$ in equation (\ref{RadReaEq}) we find that
\begin{eqnarray}\label{RadDamping}
\frac{1}{\mu}\frac{d^5{\cal D}_{ij}}{dt^5} &=& -2\frac{k\dot{r}}{r^4} \left(
9v^2-15\dot{r}^2-\frac{8k}{r}\right) \delta_{ij}
-90\frac{k\dot{r}}{r^6}(3v^2-7\dot{r}^2)r^ir^j \nonumber \\
&& + 24\frac{k}{r^5}\left(3v^2-15\dot{r}^2-\frac{k}{r}\right)(r^iv^j+r^jv^i)
+180\frac{k\dot{r}}{r^4}v^iv^j,
\end{eqnarray}
which is the approximate expression to be used in the radiative damping term 
\begin{equation}\label{RadRea2A}
{\cal R}^i = \frac{2 G_0}{15 c^5} \frac{d^5{\cal D}_{ij}}{dt^5} r^j
\end{equation}
in order to obtain equation (\ref{RadiationReaction}).  In deriving 
equation~(\ref{RadDamping}), 
we have repeatedly used the fact that 
${\bf r}\cdot{\bf v} = r\dot{r}$. 
In the following sections, we shall use equations~(\ref{BasicEq})
and~(\ref{RadiationReaction})
to study the dynamical behavior of the Kepler system over a long period of
time.

Let us now return to equation~(\ref{RadReaEq}) and point out that this equation
of relative motion --- in which the effects of emission and
absorption of gravitational radiation have been taken into account
in the quadrupole approximation and all other retardation effects have
been neglected --- is consistent with the conservation laws of energy and
momentum. We will illustrate this point explicitly for the energy of the system;
the case of angular momentum is similar but will not be treated here.
Let us therefore derive an energy expression for the equation of relative
motion (\ref{RadReaEq}). Multiplying both sides of that equation by 
the reduced mass and the relative velocity results in
\begin{equation}\label{RelEng1}
\frac{d}{dt}\left[ E + \frac{G_0}{45 c^5}
\left(\stackrel{....}{\cal D}_{ij}\dot{\cal D}^{ij}-
   \stackrel{...}{\cal D}_{ij}\ddot{\cal D}^{ij}\right)\right] =  
   -\frac{G_0}{45 c^5}
\stackrel{...}{\cal D}_{ij}\stackrel{...}{\cal D}^{ij}
-\frac{1}{6}\epsilon{\cal K}_{ij}\dot{\cal D}^{ij},
\end{equation}
where we have used 
$\dot{\cal D}_{ij} = 
  \mu\left[3(v^ir^j+r^iv^j)-2{\bf r}\cdot{\bf v}\:\delta_{ij}\right]$ 
and the fact that  $({\cal D}_{ij})$ 
is symmetric and traceless.  
The first term in the square brackets in
equation (\ref{RelEng1}) is the Keplerian energy of the relative orbit 
$E$, $2E=\mu v^2-2k\mu/r$,
and the second term is a relativistic contribution.  The rate at which this latter
term varies averages out --- over the period of the unperturbed Keplerian
motion ---  to a quantity that is negligible at the level of approximation 
under consideration in this paper.  It
follows that the average rate of loss of Keplerian energy of the dynamical
system by radiation damping is equal to the average rate at which energy leaves
the orbit via gravitational waves, as expected.  This latter rate is given by
averaging equation (\ref{EnergyEq}) over one period of elliptical Keplerian
motion and the result is \cite{peters}
\begin{equation}
\left<\frac{d{\cal E}}{dt}\right> =
\frac{32G_0^4m_1^2m_2^2(m_1+m_2)}{5c^5a^5(1-e^2)^{7/2}}
\left(1+\frac{73}{24}e^2+ \frac{37}{96}e^4\right),
\end{equation}
where $a$ and $e$ are the semimajor axis and the eccentricity of the Keplerian
ellipse, respectively.  At this average rate, gravitational radiation energy
permanently leaves the system and goes off to infinity.  No energy is actually
lost; the orbital energy is simply converted into gravitational radiation
energy.

It is sometimes stated (cf. \cite{ww} and references therein) that the force
due to gravitational radiation damping could be derived from a quadrupole
potential.  If so, the dynamical system under consideration here would be
Hamiltonian.  That is, a simple comparison of the two perturbing influences in
equation (\ref{RadReaEq}) reveals that this system would be Hamiltonian if
${d^5{\cal D}_{ij}}/{dt^5}$ were simply a function of time.  However, this
cannot be the case since
${d^5{\cal D}_{ij}}/{dt^5}$ is  a function of ${\bf r}$ and ${\bf v}$;
moreover, the system is manifestly dissipative.
\section{Continuation of Periodic Orbits}\label{conpo}
The investigation of the dynamics of the perturbed Kepler  system
(\ref{BasicEq}) can be simplified considerably if the equations of relative
motion are expressed in terms of action-angle variables that are suited to the
unperturbed system \cite{Kov,Stern}.  If the perturbations of the Kepler system
(\ref{MathEQM}) are turned off at a given instant, the relative motion will
follow an osculating elliptical orbit with semimajor axis $a$, eccentricity $e$,
eccentric anomaly $\hat{u}$, and true anomaly $\hat{v}$.  
The orientation of the underlying coordinate system can be so chosen that
the relative orbital angular momentum is positive. The action-angle
variables appropriate to the Kepler problem are the Delaunay elements
\cite{Kov,Stern} that are given for the planar problem under consideration by
\begin{eqnarray}\label{PProblem}
L := (ka)^{1/2},\qquad & & G:= p_{\theta} =  L(1-e^2)^{1/2}, \nonumber \\
\ell:= \hat{u} - e \sin{\hat{u}}, \qquad & & g:=\theta - \hat{v}.
\end{eqnarray}

To analyze the dynamical behavior contained in equation (\ref{BasicEq})
in terms of Delaunay variables,  we
first put these equations in dimensionless form.  This is done in Appendix
\ref{DEQM}.  Then, the dimensionless equations of motion are expressed in terms
of Delaunay elements; this transformation is presented in Appendix
\ref{EQMDelaunay}.  This results in the following form of the dynamical
equations:
\begin{eqnarray}\label{D2EQM1}
\frac{dL}{dt} & = & -\epsilon \left(\frac{\partial{\cal C}}{\partial \ell}
\phi(t) +
\frac{\partial{\cal S}}{\partial \ell} \psi(t)\right) +  \delta f_L,
\nonumber \\
\frac{dG}{dt} & = & -\epsilon \left(\frac{\partial{\cal C}}{\partial g}
\phi(t) +
\frac{\partial{\cal S}}{\partial g} \psi(t)\right) +  \delta f_G,
\nonumber \\
\frac{d\ell}{dt} & = & \omega + \epsilon \left(\frac{\partial{\cal C}}{\partial
L} \phi(t) +
\frac{\partial{\cal S}}{\partial L} \psi(t)\right)
+  \delta f_{\ell}, \nonumber \\
\frac{dg}{dt} & = & \epsilon \left(\frac{\partial{\cal C}}{\partial G}
\phi(t) +
\frac{\partial{\cal S}}{\partial G}\psi(t)\right) +  \delta f_g,
\end{eqnarray}
where
\begin{eqnarray}\label{feq1}
f_L & = &
\frac{4}{Lr^3}\left[1-\frac{16}{3}\frac{L^2}{r} +
\left(\frac{20}{3}L^2- \frac{17}{2}G^2\right)\frac{L^2}{r^2}
+\frac{50}{3}\frac{L^4G^2}{r^3}-\frac{25}{2}\frac{L^4G^4}{r^4}\right],
\nonumber \\
f_G & = & -\frac{18 G}{L^2r^3}\left(1 -\frac{20}{9}\frac{L^2}{r}
+\frac{5}{3}\frac{L^2G^2}{r^2}\right), \nonumber \\
f_{\ell} & = & \frac{2\sin{\hat{v}}}{eL^3Gr^2}\left[4e^2+
\frac{1}{3}
\left(73G^2-40L^2\right)\frac{1}{r}
-2\left(1+\frac{70}{3}L^2-\frac{29}{2}G^2\right)\frac{G^2}{r^2}
\right.  \nonumber \\
&& \hspace*{.5in}\left.-\frac{25}{3}\frac{L^2G^4}{r^3}
+25\frac{L^2G^6}{r^4} \right],\nonumber \\
f_g & = &
-\frac{2\sin{\hat{v}}}{eL^2r^3}\left[11+
\left(7G^2-\frac{80}{3}L^2\right)\frac{1}{r}-
\frac{25}{3}\frac{L^2G^2}{r^2}+25\frac{L^2G^4}{r^3}\right],
\end{eqnarray}
and $f_L$,$f_G$, $f_{\ell}$, $f_g$ can be expressed in terms of Delaunay
elements using classical methods of celestial mechanics involving the
Bessel functions (cf.\ Appendix \ref{EQMDelaunay}).
The right hand side of the system (\ref{D2EQM1}) is  periodic
in time with frequency $\Omega$.  For $\delta = 0$, the system (\ref{D2EQM1})
has periodic orbits as proved in \cite{cmr}; moreover, this Hamiltonian system
exhibits dynamical behavior \cite{cmr,cmr1} that appears to be characteristic of
Arnold diffusion \cite{arnold,liberman}.  In the following, we show that when
radiation reaction is taken into account the periodic orbits persist for
sufficiently small radiative damping.  
For $\epsilon=0$, on the other hand, the system~(\ref{D2EQM1}) continuously  
loses energy to gravitational radiation and eventual collapse is inevitable;
moreover, this dynamical behavior is  structurally stable and is therefore expected to
persist for sufficiently small $\epsilon\ne 0$.

Let us now assume that there are
relatively prime  positive integers $m$ and  $n$ such that $m\omega = n\Omega$,
i.e. the unperturbed Keplerian orbit is in resonance with the  disturbing
function.   
Then, it turns out that
the three-dimensional resonance manifold (period manifold)
\[{\cal Z}^L = \{(L, G, \ell, g): m\omega = n\Omega\} \] is a normally
nondegenerate manifold~\cite{cmr}.
Moreover, the solution of the unperturbed system
starting at $(L_0, G_0, \ell_0, g_0)$ is given by \[t \mapsto (L_0, G_0, \omega
\bar{t} + \ell_0, g_0),\]where $\bar{t} = t - t_0$ and $t_0$ is the initial
time.  It turns out that we can set $t_0 = 0$ with no loss of generality.  It
follows from  \cite{cmr,cmr1,ccc} that we must project the  partial derivative
of the Poincar\'{e} map  with respect to $\epsilon$   onto the complement of the
range of the infinitesimal displacement.  The partial derivative of the
Poincar\'{e} map with respect to $\epsilon$ at $\epsilon = 0$ is obtained from
the solution of the second variational initial value  problem
\begin{eqnarray}\label{IVP1}
\dot{L}_{\epsilon} & = & - \left(\frac{\partial{\cal C}}{\partial \ell}
\phi(t) +
\frac{\partial{\cal S}}{\partial \ell} \psi(t)\right) + \Delta f_L,
\nonumber \\
\dot{G}_{\epsilon} & = & - \left(\frac{\partial{\cal C}}{\partial g}
\phi(t) +
\frac{\partial{\cal S}}{\partial g} \psi(t)\right) +  \Delta f_G ,
\nonumber \\
\dot{\ell}_{\epsilon} & = & -\frac{3}{L^4}L_{\epsilon}
+ \left(\frac{\partial{\cal C}}{\partial  L} \phi(t) +
\frac{\partial{\cal S}}{\partial L} \psi(t)\right)
+ \Delta f_{\ell}, \nonumber \\
\dot{g}_{\epsilon} & = &  \left(\frac{\partial{\cal C}}{\partial G}
\phi(t) +
\frac{\partial{\cal S}}{\partial G}\psi(t)\right) + \Delta f_g,
\end{eqnarray}
with
\begin{eqnarray}\label{IVP2}
L_{\epsilon}(0, L_0, G_0, \ell_0, g_0) = 0, & \hspace*{.25in} & G_{\epsilon}(0,
L_0, G_0, \ell_0, g_0) = 0, \nonumber \\
\ell_{\epsilon}(0, L_0, G_0, \ell_0, g_0) = 0, & \hspace*{.25in} & g_{\epsilon}
(0, L_0, G_0, \ell_0, g_0) = 0,
\end{eqnarray}
where  $\Delta = \delta/\epsilon$  so that
$0<\Delta<\infty$. In fact, the partial derivative is given by the vector
\[\left[L_{\epsilon}\left(m\frac{2\pi}{\Omega}, L_0, G_0, \ell_0, g_0 \right),
G_{\epsilon}\left(m\frac{2\pi}{\Omega}, L_0, G_0,\ell_0, g_0 \right),\right.
\hspace*{1in}\]
\[ \hspace*{1in}\left. \ell_{\epsilon}\left(m\frac{2\pi}{\Omega}, L_0, G_0,
\ell_0, g_0 \right), g_{\epsilon}\left(m\frac{2\pi}{\Omega}, L_0, G_0,\ell_0,
g_0 \right)\right] .\] 
Here the independent variables
are $(t,L,G,\ell,g)$ and the last four variables of each component function
give the initial point for the original solution of the unperturbed system, i.e.
equations (\ref{D2EQM1}) with $\epsilon = 0$ and $\delta=0$.

The appropriate projection is onto the first, second and fourth coordinates.
In fact, the bifurcation function is this projection restricted to ${\cal Z}^L$,
and is given by
\begin{eqnarray}\label{BifurcationEq}
{\cal B} & = & \left[L_{\epsilon}\left(m\frac{2\pi}{\Omega}, L_0, G_0, \ell_0,
g_0 \right), G_{\epsilon}\left(m\frac{2\pi}{\Omega}, L_0, G_0,\ell_0,
g_0 \right), \right. \nonumber \\
& & \hspace*{.5in} \left.  g_{\epsilon}\left(m\frac{2\pi}{\Omega}, L_0, G_0,
\ell_0, g_0 \right)\right],
\end{eqnarray}
where $L=L_0$ is fixed by the choice of resonance, i.e. 
$m/L_0^3 = n \Omega,$ and $G_0>0$ by our choice of spatial coordinate axes.
Dropping subscripts indicating initial values of Delaunay elements for the 
sake of simplicity, 
the components of the bifurcation function (\ref{BifurcationEq}) can now be
written as
\begin{eqnarray}\label{BifrucationExp}
{\cal B}^L(G, \ell, g) & = & -\frac{\partial {\cal I}}{\partial \ell} 
                  +\Delta\int_0^{2\pi m/\Omega}f_L(L,G,\omega t+\ell,g)\,dt,
\nonumber \\
{\cal B}^G(G, \ell, g) & = & -\frac{\partial {\cal I}}{\partial g} 
                  +\Delta\int_0^{2\pi m/\Omega}f_G(L,G,\omega t+\ell,g)\,dt,
\nonumber \\
{\cal B}^g(G, \ell, g) & = & \frac{\partial {\cal I}}{\partial G} 
                  +\Delta\int_0^{2\pi m/\Omega}f_g(L,G,\omega t+\ell,g)\,dt,
\end{eqnarray}
where ${\cal I}$ is defined by
\begin{equation}\label{IEq1}
{\cal I} = \int_0^{2\pi m/\Omega} [{\cal C}(L, G, \omega t + \ell, g)\phi(t) +
{\cal S}(L, G, \omega t + \ell, g)\psi(t)]\: dt,
\end{equation}
and $\cal C$ and $\cal S$ are given in Appendix~\ref{EQMDelaunay}.
The integral in (\ref{IEq1}) has been calculated in \cite{cmr} with the
result that ${\cal I} = 0$ for $n > 1$, while for $n = 1$,
\begin{eqnarray}\label{Ieqn3}
{\cal I} & = & \frac{1}{2}\pi ma^2\Omega \{ \alpha(A_m \cos{m \ell}
\cos{2g} - B_m \sin{m \ell} \sin{2g}) \hspace*{1in} \nonumber \\
& & \hspace*{.5in} + \beta [A_m
 \cos{(m\ell - \rho)} \sin{2g} + B_m
\sin{(m\ell- \rho)} \cos{2g}] \}.
\end{eqnarray}
The periodic orbits that continue are determined by the simple zeros of the
bifurcation function (\ref{BifrucationExp}).  The integrals multiplying
$\Delta$ in equations (\ref{BifrucationExp}) are given in
Appendix \ref{RadForceApp}; therefore, 
periodic orbits continue to exist when the following equations are
satisfied by simple zeros:
\begin{eqnarray}\label{ZerosBif1}
\frac{\partial {\cal I}}{\partial \ell} + 2\pi \Delta \frac{L^3}{G^7}\left( 8 +
\frac{73}{3} e^2 + \frac{37}{12} e^4 \right) & = & 0, \nonumber \\
\frac{\partial {\cal I}}{\partial g} + 2\pi \Delta \frac{1}{G^4} ( 8 +
7 e^2) & = & 0, \nonumber \\
\frac{\partial {\cal I}}{\partial G}  & = & 0.
\end{eqnarray}
For $n > 1$, ${\cal I} = 0$ and hence these equations have no
solution as the expressions multiplying $\Delta$ are manifestly positive.
For $n = 1$  and $\Delta = 0$, we have proved in \cite{cmr} that there are
simple zeros of the bifurcation function for all positive integers $m$.  It
follows that equation (\ref{ZerosBif1}) must have simple zeros as well for
sufficiently small $\Delta$.

It is important to remark here that this conclusion would not be altered
by the inclusion of terms of order $\epsilon^2$, $\epsilon\delta$, 
$\delta^2$, or higher in the original system (\ref{BasicEq}). In fact,
the incorporation of such higher order effects could only affect the shape
of a periodic orbit of the type investigated here but not its existence.

The Kepler system under investigation in this paper may in general
have other periodic orbits than those revealed by our first order method.
The investigation of the stability of the periodic orbits under consideration
is beyond the scope of this work. If
there is a stable periodic orbit for $\epsilon>0$ and $\delta>0$,
then its (open) basin of attraction consists of trajectories
that are permanently  captured into resonance.
This same
issue will be discussed in connection with the averaging methods that
we introduce in the next section.
\section{Averaged Dynamics}\label{avedyn}
In this section, we wish to consider our system in Delaunay elements
when no external gravitational waves are present.  
The motivation for this discussion is
the experimental observation of the inward spiraling of the members of the
Hulse-Taylor binary pulsar \cite{hulse}.  
Thus we consider the traditional Kepler problem except that the
post-Newtonian effect of gravitational radiation damping is taken into
account in the quadrupole approximation. We expect that the system
remains planar and loses energy and angular momentum
such that eventually $L\to 0$, $G \to 0$, and the system collapses.
The behavior of the system in the infinite past is less trivial.
The precise manner in which the orbit behaves on average for 
$t\to\pm\infty$ can be determined using the method of averaging.
In this case, we have
\begin{eqnarray}\label{HulseEQM}
\dot{L} =  \delta f_L, & &\qquad \dot{G} = \delta f_G, \nonumber \\
\dot{\ell} = \omega +  \delta f_{\ell}, & & \qquad \dot{g} = \delta f_g.
\end{eqnarray}
Equations (\ref{HulseEQM}) are of the general form
\begin{equation}\label{HulseEQM1}
\dot{I} = \epsilon f(I, \varphi, \epsilon), \qquad 
\dot{\varphi} = \omega(I) + \epsilon g(I, \varphi, \epsilon),
\end{equation}
where both $f$ and $g$ are $2\pi$ periodic in the angle variable $\varphi$.
In fact, in our case $I = (L, G, g)$ and $\varphi = \ell$.  The
Averaging Theorem~\cite{arnold2} asserts the following:  
Suppose that $t \mapsto (I(t), \varphi(t))$ 
is a solution of equations (\ref{HulseEQM1}) and $\omega
(I(t)) > 0$ for all $t$.  If $t \mapsto J(t)$ is the solution of the averaged
system
\[ \dot{J} = \epsilon \frac{1}{2\pi} \int_0^{2\pi} f(J, \varphi, 0) d\varphi \]
with $J(0) = I(0)$, then, for sufficiently small $\epsilon$, there is a constant
$C$ independent of $\epsilon$ such that
\[
|I(t) - J(t)| < C\epsilon \qquad \mbox{\rm for} \qquad 
    0 \leq t \leq \frac{1}{\epsilon}.
\] 
That is, the ``action variables" of the original
system (\ref{HulseEQM1}) remain close to the solution of the averaged system
over a long time-scale.

It is important to note that the Averaging Theorem, in the generality stated
above, is only true for the case of a single frequency, i.e. there is only one
angle variable and its frequency does not vanish.  If there are more than one
angle present, then resonances among the frequencies must be taken into account.
Though $g$ is an angle variable, its evolution is slow and
its average rate of variation vanishes;
hence, it may be regarded as an ``action'' variable for the purposes of this
section. Therefore in our case,
the averaged system obtained from (\ref{HulseEQM}), where we use
$L$, $G$, and $g$ for the averaged variables,
$e = \sqrt{L^2-G^2}/L$, and $\Delta=\delta/\epsilon$, 
is given by (cf. Appendix~\ref{RadForceApp})
\begin{eqnarray}\label{AveEQM}
\dot{L} & = & -\epsilon \frac{\Delta}{G^7} \left(8 + \frac{73}{3}e^2 +
\frac{37}{12}e^4\right) , \nonumber \\
\dot{G} & = & -\epsilon \frac{\Delta}{L^3G^4}(8 + 7 e^2), \nonumber \\
\dot{g} & = & 0.
\end{eqnarray}

To determine the dynamics of the averaged system (\ref{AveEQM}) for the
variables $L$ and $G$, we note that equations (\ref{AveEQM}) are autonomous.  
Of course, the system is in this case  singular on the (invariant)
$L$-axis. Also, recall that it suffices to consider only  the case where
$G>0$. After multiplication by
$12L^5G^7/\epsilon\Delta$, we obtain the dynamically equivalent system of
equations
\begin{eqnarray}\label{AveEQM2}
\dot{L} & = & -L(37G^4-366G^2L^2+425L^4), \nonumber \\
\dot{G} & = & -G^3(180L^2-84G^2).
\end{eqnarray}
The Delaunay elements are defined only for $L \geq 0$ and  
$0<G \leq L$. 
It is easy to determine the phase portrait of the system~(\ref{AveEQM2});
in fact, all orbits in the sector bounded by the $L$-axis and the line 
$L = G$ are attracted to the origin.  
The ``singular set" corresponding to the limits of the definition of
Delaunay elements, $e=0$ and $e=1$, consists of the line
$L=G$ and the $L$-axis, respectively; indeed, these are invariant
sets in the scaled system~(\ref{AveEQM2}). 
Recall that $L = \sqrt{a}$ 
and $G=\sqrt{a(1-e^2)}$, where $a$ and $e$ are the semimajor axis and the
eccentricity, respectively.  From our analysis,  
it follows that $a$ approaches
zero as $t\to\infty$  and $a$ increases without bound as 
$t \rightarrow - \infty$.  
We claim that $e \rightarrow 0$ as $t \rightarrow \infty$ and 
$e \rightarrow 1$ as $t \rightarrow - \infty$.  
To see this just note that
\[ 
\frac{d}{dt}\frac{G}{L} = \frac{G}{L}(L^2-G^2)(425L^2-121G^2);
\] 
now let $\eta = G/L$ and observe that $0 < \eta < 1$, 
$\eta=\sqrt{1-e^2}$,
and
\[\dot{\eta} = L^4\eta(1-\eta^2)(425-121\eta^2).\]
By rescaling time, the dynamics of this equation turns
out to be equivalent to the dynamics of
\[\eta' = \eta(1-\eta)(425-121\eta^2),\]
where a prime denotes differentiation with respect to the new temporal variable.  
The last differential equation has a source at $\eta=0$, 
a sink at $\eta=1$, and no other rest point on the interval
$(0,1)$ on the corresponding phase line; therefore,  
we see that $\eta \rightarrow 1$ as $t\to\infty$ and $\eta \rightarrow 0$ 
as $t\to -\infty$, i.e. 
$e \rightarrow 0$ as $t\to\infty$  and $e \rightarrow 1$ 
as $t\to-\infty$.
Hence the orbit is unbound in the infinite past.

The main physical conclusion of this section has been previously obtained
by Walker and Will~\cite{ww}. However, the method of averaging employed here
provides a simple and
transparent proof of their results on the dissipative
nature of the radiation damping.  
\section{Passage Through Resonance}
In this section, we consider some additional aspects of the dynamics of
equation~(\ref{BasicEq}) that can be determined using the method of averaging.
In particular, we will discuss the phenomenon of capture into resonance. 

The Delaunay ``action-angle'' 
variables provide the appropriate form of the dynamical
equations required for the averaging method. Here we will begin with
system~(\ref{D2EQM1}) in the form
\begin{eqnarray}\label{AAAveraging}
\dot{L} & = & -\epsilon \frac{\partial {\cal H}^*}{\partial \ell} +
\epsilon\Delta f_L, \nonumber \\
\dot{G} & = & -\epsilon \frac{\partial {\cal H}^*}{\partial g} +
\epsilon\Delta f_G \nonumber \\
\dot{g} & =  & \quad\epsilon \frac{\partial {\cal H}^*}{\partial G} +
\epsilon\Delta f_g, \nonumber \\
\dot{\ell}& = &\quad \frac{1}{L^3} + \epsilon \frac{\partial  {\cal H}^*}{\partial
L} + \epsilon\Delta f_{\ell}, \nonumber \\
\dot{s} & = & \quad \Omega,
\end{eqnarray}
where 
\[
{\cal H}^*=\frac{1}{2}\Omega^2\left[
\alpha {\cal C}(L,G,\ell,g)\cos s+\beta {\cal S}(L,G,\ell,g)\cos(s+\rho)
\right],
\] 
denotes the $O(\epsilon)$ terms of the Hamiltonian giving
the gravitational wave interaction in Delaunay elements and $s := \Omega t$.
Here we assume that $\epsilon$ is the small parameter; however, our
conclusions remain the same if $\delta=\epsilon\Delta$ is taken to be the
small parameter in this system.

A key observation that motivates the analysis of this section is the fact
that the angular variable $g$ is a slow variable for~(\ref{AAAveraging}).
Thus, the dynamical system can be treated as a two-frequency system with
fast variables $\ell$ and $s$.  The associated 
resonances are given by relations between the frequencies $1/L^3$ and
$\Omega$.   More precisely, if $m$ and $n$ are relatively prime integers,
the $(m:n)$ resonant manifold is given by
\[\{(L,G,g,\ell,s): m\frac{1}{L^3} = n \Omega\}. \] 

After averaging system~(\ref{AAAveraging})
over the fast variables, we obtain the system~(\ref{AveEQM}) that was analyzed
in the last section. Thus, according to the Averaging Principle~\cite{arnold2},
every perturbed Keplerian elliptical orbit collapses even in the presence of
gravitational wave interaction.
However, as is well known, the Averaging
Principle is not valid for two-frequency systems unless additional hypotheses
are imposed.  Violations of the asymptotic estimates
embodied in the Averaging Principle are associated with the dynamical phenomenon
called  capture into resonance. 
A theorem of A.\ I.~Neishtadt (cf. \cite{arnold2}, p.~163, for a
discussion) allows for the possibility of 
capture into resonance and provides a strong result on
the applicability of the Averaging Principle.

Neishtadt's Theorem requires two hypotheses that we refer to as condition
$N$ and condition $B$.  Condition $N$ requires
that the rate of change of the frequency ratio of the
fast angles with respect to
time along the {\it averaged} system be bounded away from zero.  
For our case, the frequency ratio is given by $(1/L^3)/\Omega$, and the
required derivative is 
\begin{equation}
\frac{3\epsilon\Delta }{\Omega L^4 G^7}\left( 8 + \frac{73}{3} e^2 +
\frac{37}{12} e^4\right).
\end{equation}
Clearly, the derivative is bounded away from zero as long as 
$L$ and $G$ are bounded away from infinity.  
As we have shown in Section~\ref{avedyn}, both $L$ and $G$ decrease with time in the
averaged motion.  Thus, condition $N$ is satisfied along each fixed
orbit of the averaged system.
Condition $B$ is generically satisfied for most orbits.

If conditions $N$ and $B$ both hold, then
Neishtadt's Theorem asserts that for all initial points outside a set
of measure not exceeding a constant multiple of $\sqrt{\epsilon}$, the averaged
motion approximates the motion over a time-scale $1/\epsilon$ with
an error given by $O(|\ln{\epsilon}|\sqrt{\epsilon})$.  In particular, capture
into resonance is rare, and most motions are well approximated by the averaged
motion. This is the behavior predicted for our model equations. 
That is, except for a set of initial conditions with small measure, 
the variables $L$ and $G$ associated with the perturbed
orbit are such that $L\to 0$ and $G\to 0$ as $t\to\infty$; 
the long time behavior of
the osculating ellipse is described by eventual collapse.

The exceptional set of initial conditions 
mentioned in Neishtadt's Theorem is not empty in our case.
In fact, we have proved that some of the unperturbed periodic orbits 
can be continued to periodic orbits in the presence of perturbation
for sufficiently small $\epsilon$ and $\Delta$; indeed, 
these periodic orbits are in a sense
permanently captured into resonance. 

In order to determine the average dynamics in
more detail, the dynamical behavior near the resonant manifolds must be studied.
One approach to the study of this behavior is provided by the theory of
partial averaging near a resonant manifold. In fact,
these ideas are used in the proof of Neishtadt's Theorem. 
An elementary exposition of the main aspects of the now well established theory 
can be found, for example, in \cite{arnold2}; a mathematical proof of
Neishtadt's Theorem is given in~\cite{lochak}. 
This theory requires second order averaging near the resonant manifold,
where we find essentially a pendulum-like equation with slowly varying
amplitude, phase, damping, and torque. The existence and stability
of its stationary solutions together with the positions of the
associated invariant manifolds determine the actual average dynamics near
each resonance.
We have obtained the second order partially averaged system for
the perturbed Kepler problem, but its analysis
is beyond the scope of this paper. However, we mention here two facts
in this connection: Near $(m:n)$ resonance with $n\ne 1$, there is no
capture into resonance; moreover, as $\Delta$ increases, the likelihood
of capture into resonance decreases. 

Figure~\ref{capfig} shows an example of capture into resonance.
The top panel in this figure illustrates capture into resonance and
the middle panel involves an orbit whose initial conditions are slightly
perturbed compared to the orbit depicted in the top panel.
The middle panel shows passage through resonance on the
time-scale of our numerical experiment.
In the bottom panel, the behavior of the orbital angular
momentum $G$ is plotted versus $L=\sqrt{a}$ for
the system depicted in the middle panel.
During this passage through  $(1:1)$ resonance, the energy of the orbit
is constant on average while the orbital angular momentum increases in
this case --- i.e. the eccentricity of the osculating ellipse decreases
from $e\approx 0.8$ to $e\approx 0.4$. That is, there is  balance
on the average between the processes of emission and absorption
for energy but not for angular momentum.  Recall that the resonance 
condition only fixes the energy of the osculating 
orbit and so the angular momentum can change.
A preliminary analysis indicates that this behavior is consistent with
the second order partially averaged dynamics; however, a full investigation of this
interesting phenomenon necessitates further research. 
It would be quite interesting, of course, if this theoretically
rare phenomenon could be observed astronomically:
A binary system gradually spirals inward, but when the decreasing
semimajor axis reaches a certain value corresponding to resonance
with an external periodic perturbation the collapse process
temporarily ceases --- though the binary orbit's eccentricity
could change considerably in this sojourn while the semimajor axis 
fluctuates with increasing amplitude about the resonance value ---
during the period of passage through resonance that may be very long
compared to the orbital period until the collapse process
resumes again.
We also note that
for our choice of $\epsilon$ the time-scale for the validity of the
averaged motion for orbits not captured into resonance is $1/\epsilon=10^4$,
a value that is an order of magnitude smaller than the integration time of
$10^5$. Finally, we emphasize that the numerical experiments are conducted
by integrating the original equations of motion --- not the averaged system.
\section{Circularly Polarized Forcing Function}
A remarkable outcome of our nonlinear analysis of  {\it gravitational
ionization} \cite{cmr} has been the recognition that for normally incident
monochromatic plane waves with definite helicity (i.e. right or left circularly
polarized waves) the KAM theorem would ensure that there would be no ionization
for $\epsilon < \epsilon_{\mbox{\rm \tiny KAM}}$.  This result has been further
discussed in \cite{cmr1}.  It is clear on physical grounds that the inclusion of
radiative dissipation cannot change the basic result of this analysis:  The
system would constantly lose energy and final collapse would be inevitable.
While the KAM theorem only guarantees confinement, damping by radiation cannot
but lead to catastrophic collapse.  For $\epsilon > \epsilon_{\mbox{\rm \tiny
KAM}}$, the situation is less clear and --- as discussed in previous sections
--- there could be an interesting interplay between collapse and Arnold
diffusion.

It must be pointed out that these results are independent of the frequency of
incident radiation $\Omega$ and only require that $\alpha = \beta$ and $\rho =
\pm\pi/2$.  The dynamical system can be viewed from a frame rotating with
frequency $\Omega /2$; in this frame, there exists an energy integral once the
radiative damping is neglected.  That is, the incident gravitational wave stands
completely still as a consequence of the gravitational helicity-rotation
coupling
\cite{mashhoon3} and hence the dynamical system is time independent in the
absence of damping.  The KAM theorem would then imply that for sufficiently
small $\epsilon$  the system is forever bounded.  This confinement brings into
rather sharp focus the nonreciprocal nature of emission and absorption of
gravitational radiation.  Clearly, the continuous loss of energy by the binary
would eventually result in the collapse of the system.
\section{Transient Chaos}
The equations of motion~(\ref{MathEQM}) can be integrated numerically given a set
of initial conditions at $t = 0$.  For this purpose it is convenient to 
express these equations in dimensionless form using the 
scale transformations discussed in Appendix~\ref{DEQM}. The actual system that
we use is given by 
\begin{eqnarray}\label{NumExpEq}
\frac{dr}{dt}& = &p_r, 
    \qquad  \frac{d\theta}{dt} = \frac{p_{\theta}}{r^2},\nonumber\\
\frac{dp_r}{dt} & = & -\frac{1}{r^2} + \frac{p_{\theta}^2}{r^3}+4\delta
\frac{p_r}{r^3}\left(p_r^2+6\frac{p_{\theta}^2}{r^2}+\frac{4}{3r} \right)
\nonumber \\
& & \hspace*{.25in}  -\epsilon
r\Omega^2[\alpha\cos{2\theta}\cos{\Omega t} + \beta\sin{2\theta}\cos{(\Omega t +
\rho)}], \nonumber \\
\frac{dp_{\theta}}{dt} & = & 2\delta \frac{p_{\theta}}{r^3}\left(9 p_r^2 -
6\frac{p_{\theta}^2}{r^2}+\frac{2}{r}\right) \nonumber \\
& & \hspace*{.25in}
+\epsilon r^2\Omega^2[\alpha\sin{2\theta}\cos{\Omega t} -
\beta\cos{2\theta}\cos{(\Omega t + \rho)}].
\end{eqnarray}
We remark that this system is numerically ill-conditioned near collision, 
i.e. near $r=0$. 
Unfortunately, there does not seem to be a way to regularize the system
for numerical integration as can be done for regular perturbations of Kepler motion
as discussed in~\cite{Stiefel}. 
The problem is that the two perturbation terms in 
our system enter with negative and positive powers of the variable $r$. 

A possible set of initial
conditions for~(\ref{NumExpEq}) that we use for our
numerical experiments in this work 
is $(p_r, p_{\theta}, r, \theta) = (e, 1, 1, 0)$, which corresponds
to an osculating ellipse at $t = 0$ with unit orbital angular momentum,
eccentricity $e$, semimajor axis $a=(1-e^2)^{-1}$, and Keplerian frequency
$\omega = (1-e^2)^{3/2}$.  The true anomaly of this osculating ellipse is given
by $\hat{v} = \pi/2$, the Delaunay element $g = -\pi/2$, and the periastron
occurs at $\theta = -\pi/2$.
 
The main conclusion of the analysis in this paper is that
for sufficiently small $\epsilon$, 
the majority of initial conditions
lead {\em on average} to collapse. 
However, from basic results in
nonlinear  dynamics,
we expect that near resonances there would be transverse
intersections of stable and unstable manifolds 
of some of the perturbed periodic orbits. 
This suggests the existence of transient chaos --- chaotic invariant sets that
are not attractors. We will see in a moment that numerical experiments are
consistent with this observation. However, since we do not have estimates on
the size of the perturbation parameters for which the expected behavior is valid,
there does not seem to be a mathematical obstruction to the existence of
attracting chaotic sets (strange attractors) for some choices of the perturbation
parameters. In this regard, 
we mention the connection between the planar Kepler system and a 
two-dimensional anharmonic system with an 
external periodic force and damping~\cite{Stiefel}.
The existence of transient chaos for small amplitude forcing and the possibility
of the existence of strange attractors for larger amplitude 
forcing and damping is an interesting and  
well documented dynamical feature of the
systems in the latter class. In fact, 
the existence of transient chaos and strange attractors for 
the case of a single oscillator is well known (see, for example,~\cite{wig}).
It should also be mentioned that the connection between the Kepler
system and oscillators is quite general; in this paper, we have restricted our
attention to the planar Kepler problem for the sake of simplicity.  It is
possible that the three-dimensional Kepler system given by the
general form of  equations
(\ref{BasicEq}) and (\ref{RadiationReaction}) would exhibit novel features not
found in the planar case.
We have conducted a {\it preliminary} numerical search for a strange attractor
in our perturbed Kepler system, but we have not succeeded in finding such an attractor.
This is an interesting problem for further research.

As a numerical example consistent with transient chaotic behavior,
we refer to Figure~\ref{trans} where the evolution of the action variables, 
strobed at each cycle of the incident wave, is depicted.
While both action variables tend to eventual collapse, complicated transient
dynamics is indicated in this case.
It is interesting to note that here the orbit passes through a dense set
of $(m:n)$ resonances that may be 
responsible for the complicated behavior of these
action variables. The evolution of these variables in the bottom panel should
be compared and contrasted with that in the bottom panel of Figure~\ref{capfig}.

Finally, we remark that while energy 
continuously leaves the system via the emission of gravitational waves,
the definite trend toward collapse that results may --- under special
circumstances --- be counter-balanced by the input of energy into the system by
the external periodic perturbations.  If the net flow of energy into the system
is positive on average, gravitational ionization will take place---the
semimajor axis of the osculating ellipse will grow, albeit not necessarily
monotonically. This balance seems to depend very sensitively
on the choice of parameters and initial conditions. For
instance, let the initial conditions be 
$(p_r, p_{\theta}, r, \theta) = (0.5, 1, 1, 0)$ in the system~(\ref{NumExpEq})
and set the parameters such that
$\alpha = 2$, $\beta = 2$, $\rho = 0$,
and $\Omega = 1.299038106$;  for $\epsilon = 0.005$ and 
$\delta/\epsilon = 0.00198992$ the
system appears to ionize over a long time-scale, 
while for $\delta/\epsilon = 0.00198993$ the system collapses.
\appendix
\section{Scale Transformations}\label{DEQM}
To facilitate the analysis of the dynamical behavior of the Kepler
problem with both incident gravitational waves and gravitational radiation
damping, we transform the equations of motion for this system to dimensionless
form.   To this end, let $r^i = R_0 \hat{r}^i$ and $t = T_0 \hat{t}$, where
$R_0$ and $T_0$ are scale parameters and $\hat{r}^i$ and $\hat{t}$ are
dimensionless.  This will be true for all quantities with a hat.  Substituting
for $r^i$ and $t$ in equations~(\ref{BasicEq}) and~(\ref{RadiationReaction}) 
leads to
\begin{eqnarray}\label{HatEqOfMotion}
\frac{d^2\hat{r}^i}{d\hat{t}^2} +
\hat{k}\frac{\hat{r}^i}{\hat{r}^3} +
\frac{\delta}{\hat{r}^3}\left[\left(12
\hat{v}^2  -30\dot{\hat{r}}^2-
4\frac{\hat{k}}{\hat{r}}\right)\frac{d\hat{r}^i}{d\hat{t}} -
\frac{\dot{\hat{r}}} {\hat{r}}\left(36\hat{v}^2 \right. \right. & &
\nonumber \\
\left. \left. -50\dot{\hat{r}}^2 + \frac{4}{3}\frac{\hat{k}}{\hat{r}} \right)
\hat{r}^i\right] + \epsilon \hat{{\cal K}}_{ij}\hat{r}^j   & = & 0,
\end{eqnarray}
where
\begin{displaymath}
\hat{k} := k\frac{T_0^2}{R_0^3}, \qquad \delta := \frac{4}{5}
\frac{G_0^2m_1m_2}{c^5T_0R_0}, 
\qquad \mbox{\rm and} \qquad  
\hat{\cal K}_{ij} :=
T_0^2{\cal K}_{ij}.
\end{displaymath}
Let us now consider the physical interpretation of such a transformation.  
If we take a
binary system with initial period $2\pi T_0$ and semimajor axis $R_0$, then by
Kepler's law
\[
\omega^2 = \frac{G_0(m_1+m_2)}{R_0^3} = \frac{1}{T_0^2},
\]
i.e. $\hat{k} = 1$.
In general, one can define $\hat{\omega}=\omega T_0$ and  $\hat{a}=a/R_0$;
then, Kepler's law becomes $\hat{\omega}^2=\hat{k}/\hat{a}^3$, where $\hat{a}$
is the dimensionless semimajor axis.  We also note that $\delta$ is
dimensionless and turns out to be a small parameter, $0<\delta<<1$, for all
realistic binary systems.  By setting
$\hat{k} = 1$ and dropping all hats in equation (\ref{HatEqOfMotion}), we obtain
\begin{eqnarray}\label{FinalEqOfMotion}
\frac{d^2 r^i}{dt^2} + \frac{r^i}{r^3} + \frac{\delta}{r^3} \left[\left(12v^2
-30\dot{r}^2-\frac{4}{r}\right)\frac{dr^i}{dt} -\frac{\dot{r}}{r}\left(36v^2
\right. \right. & &    \nonumber \\
\left. \left. -50\dot{r}^2 + \frac{4}{3r} \right)r^i\right] +
\epsilon {\cal K}_{ij}r^j   & = & 0,
\end{eqnarray}
which is the form of the equation of motion that we investigate in this paper.
Now everything is dimensionless in equation (\ref{FinalEqOfMotion}); there are
two small parameters $\epsilon$ and
$\delta$ and because $\hat{k}=1$, the underlying scales $R_0$ and $T_0$ are
connected such that $R_0^3 = k T_0^2$.  Furthermore, if we take the 
direction of incidence of the external
gravitational waves to be perpendicular to the initial
orbital plane, then it follows
directly from~(\ref{FinalEqOfMotion}) that the damped motion
is planar.

Let us now estimate the magnitude of $\delta$ for binaries of physical
interest.  Expressing $\delta$ in terms of the total mass $M = m_1+m_2$ and the
reduced mass $\mu = m_1m_2/M$, we find
\[\delta =
\frac{4}{5}\left(\frac{\mu}{M}\right)\left(\frac{R_0}{cT_0}\right)\left(
\frac{G_0M}{c^2R_0}\right)^2.\]
Furthermore, our assumption that $\hat{k}=1$ implies
\[\frac{R_0}{cT_0} = \left(\frac{G_0M}{c^2R_0}\right)^{1/2};\]
hence, we have
\[\delta =
\frac{1}{5}\left(\frac{4\mu}{M}\right)\left(\frac{G_0M}{c^2R_0}
\right)^{5/2}.\]
Here $4\mu/M \leq 1$ by definition, and $G_0M/c^2R_0$ is always less than
$1/2$.  The quantity $2G_0M/c^2R_0$ is the ratio of the Schwarzschild radius of
the system to the semimajor axis of the binary; thus, the system is nearly a
black hole for $2G_0M/c^2R_0 \sim 1$, while for a Keplerian binary
$2G_0M/c^2R_0<<1$, which indicates that the semimajor axis of the orbit is much
larger than the gravitational radius of the binary system.  It follows from
these remarks that for a physical system $\delta < 0.04$.  For the Earth-Sun
system, for instance, $\mu/M \simeq 3 \times 10^{-6}$ and $G_0M/c^2R_0 \simeq
10^{-8}$, so that $\delta \simeq 10^{-26}$; by comparison, $\delta$ is
negligibly small for an artificial satellite in orbit about the Earth.  On
the other hand, for the Hulse-Taylor binary pulsar, PSR 1913+16, we find that
$\delta \simeq 10^{-15}$.
\section{Equations of Motion in Delaunay \newline Elements}\label{EQMDelaunay}
Let us now
transform the equations of motion (\ref{FinalEqOfMotion}) from Cartesian
coordinates and velocities to Delaunay elements.
The standard derivation of the equations of motion in Delaunay variables
is based on the assumption that the system under consideration is Hamiltonian.
However, the dynamical system in our treatment is dissipative; therefore,
in this appendix we present a general derivation for the planar Kepler problem.
This transformation  is performed in two steps.  First
equation (\ref{FinalEqOfMotion}) is expressed in polar coordinates and then
in Delaunay elements.  Let
\[
f_i =
-\frac{1}{r^3}\left[\left(12v^2-30\dot{r}^2-\frac{4}{r}\right)
\frac{dr^i}{dt}
-\frac{\dot{r}}{r}\left(36v^2-50\dot{r}^2+\frac{4}{3r}\right)r^i
\right] 
\]
and ${\bf r} = (r\cos{\theta}, r\sin{\theta}, 0)$; then,
equation (\ref{FinalEqOfMotion}) can be written in polar coordinates as
\begin{eqnarray}\label{PolarEQM}
\ddot{r} - r\dot{\theta}^2 + \frac{1}{r^2} &=&\delta f_r - \epsilon {\cal
K}_{r}, \nonumber \\
r\ddot{\theta} +2\dot{r}\dot{\theta} & = & \delta f_{\theta} - \epsilon {\cal
K}_{\theta},
\end{eqnarray}
where
\begin{displaymath}
f_r = f_1 \cos{\theta} + f_2 \sin{\theta},\qquad
f_{\theta} = -f_1 \sin{\theta} + f_2 \cos{\theta},
\end{displaymath}
\begin{equation}\label{frftheta}
{\cal K}_r = r({\cal K}_{11}\cos{2\theta}+{\cal K}_{12}\sin{2\theta}),
\qquad 
{\cal K}_{\theta} = r(-{\cal K}_{11}\sin{2\theta}+{\cal K}_{12} \cos{2\theta}).
\end{equation}
In determining ${\cal K}_{r}$ and ${\cal K}_{\theta}$,  we have utilized the
fact that the matrix $({\cal K}_{ij})$ is symmetric and  traceless.  In what
follows, we let
\begin{equation}\label{Hatdefn}
F_r: = \delta f_r - \epsilon {\cal K}_{r},\qquad
F_{\theta}:= \delta f_{\theta} - \epsilon {\cal K}_{\theta}
\end{equation}
for the sake of simplicity.

The Delaunay elements are defined by the ellipse tangent to the path at time $t$.
The state of relative motion is specified by $(r,\theta,p_r,p_\theta)$ in
polar coordinates; therefore, using the standard formulae for elliptic motion
we have $p_rp_\theta=e\sin\hat{v}$ and $p_\theta^2/r=1+e\cos\hat {v}$,
which uniquely specify the eccentricity $e$ and the true anomaly $\hat{v}$ of
the osculating ellipse. Next, the semimajor axis of this ellipse $a$ is 
obtained from $p_\theta^2=a(1-e^2)$ and
the eccentric anomaly $\hat {u}$ can be determined from 
\begin{displaymath}
\cos{\hat{v}} = \frac{\cos{\hat{u}}-e}{1-e\cos{\hat{u}}}\:, \qquad
\sin{\hat{v}}=\frac{\sqrt{1-e^2}\sin{\hat{u}}}{1-e\cos{\hat{u}}}.
\end{displaymath}
The Delaunay elements are then given by
\begin{eqnarray}\label{DelaunayElements}
L = \sqrt{a},\: & & \: G = p_{\theta}, \nonumber \\
\ell =\hat{u}-e\sin{\hat{u}},\: & & \: g=\theta-\hat{v}.
\end{eqnarray}
To obtain the equations of motion in Delaunay elements, we observe that the
energy and orbital angular momentum of the relative motion per unit mass are
\begin{eqnarray}\label{EllipseEnergy}
\frac{E}{\mu} & = & \frac{1}{2}(\dot{r}^2+r^2\dot{\theta}^2)-\frac{1}{r} = -\frac{1}{2a},
\nonumber \\
p_{\theta} & = & r^2\dot \theta= \sqrt{ a(1-e^2)},
\end{eqnarray}
where the orientation of the spatial coordinate system is chosen such
that $p_\theta>0.$
Now
\begin{displaymath}
\frac{1}{\mu}\frac{dE}{dt} = {\bf F} \cdot {\bf v},
\end{displaymath}
where
$
{\bf F}\cdot {\bf v} = F_r \: \dot{r} + F_{\theta} \:
r\dot{\theta}
$.
It follows that
\begin{equation}\label{Leqn}
\frac{dL}{dt}  = \frac{a}{\sqrt{1-e^2}} \left[F_r e \sin{\hat{v}} 
                 + F_{\theta} \frac{a(1-e^2)}{r}\right],
\end{equation}
where we have replaced $\dot{r}$ by $e \sin{\hat{v}}/p_{\theta}$, which follows
from  equation (\ref{EllipseEnergy}). It is a direct consequence of equations
(\ref{PolarEQM}) and (\ref{Hatdefn}) that
\begin{equation}\label{Geqn}
\frac{dG}{dt} = r F_{\theta}.
\end{equation}
To determine $d\ell/dt$ and $dg/dt$, we first note the following useful
relationships that are obtained from equations (\ref{Leqn}) and (\ref{Geqn}),
the definitions of the Delaunay elements, and the Kepler law $\omega = 1/L^3$,
\begin{eqnarray}\label{adotedot}
\frac{da}{dt} & = & \frac{2}{\omega\sqrt{1-e^2}}\left[F_r  \: e
 \sin{\hat{v}} +
F_{\theta} \: \frac{a(1-e^2)}{r} \right], \nonumber \\
\frac{de}{dt} & = & \frac{\sqrt{1-e^2}}{a\omega}\left[F_r  \:
\sin{\hat{v}} +
F_{\theta} \:\left(e+ \frac{r+a}{a}\cos{\hat{v}}\right) \right].
\end{eqnarray}
Consider first $dg/dt = d\theta/dt - d\hat{v}/dt$, where
\begin{equation}\label{thetadot}
\frac{d\theta}{dt}  = \frac{\sqrt{a(1-e^2)}}{r^2} \: .
\end{equation}
Differentiating the relationship 
\[\ln{r} = \ln{a} + \ln{(1-e^2)} - \ln{(1+e\cos{\hat{v}})}\] 
with respect to time and using the definition of $g$ together
with equations~(\ref{adotedot}) and~(\ref{thetadot}), we have
\begin{equation}\label{geqn}
\frac{dg}{dt} = \frac{\sqrt{a(1-e^2)}}{e}\left[-F_r \: \cos{\hat{v}} +
F_{\theta}\, \Big( 1 + \frac{r}{a(1-e^2)} \Big)\sin{\hat{v}} \right].
\end{equation}
Let us now compute $d\ell/dt$ using the definition
of $\ell$: The Kepler equation $\ell=\hat{u}-e\sin{\hat {u}}$ expresses
the mean anomaly $\ell$ in terms of the eccentric anomaly and the
eccentricity of the orbit.  
Thus,
\begin{equation}\label{ell1eqn}
\frac{d\ell}{dt} =
\frac{r}{a}\: \frac{d\hat{u}}{dt}-\frac{r}{a\sqrt{1-e^2}}\sin{\hat{v}} \:
\frac{de}{dt}.
\end{equation}
Using the identity
\begin{displaymath}
\dot{r} - \dot{a}\frac{r}{a}+\dot{e}\: \frac{a-r}{e} =
\frac{er\sin{\hat{v}}}{\sqrt{1-e^2}}\: \frac{d\hat{u}}{dt},
\end{displaymath}
which is obtained by differentiating $r=a(1-e\cos{\hat{u}})$ with respect to
time, and equation (\ref{adotedot}), we find
\begin{equation}\label{elleqn}
\frac{d\ell}{dt} = \omega + \frac{r}{e\sqrt{a}}\left[F_r\:
(-2e+\cos{\hat{v}}+e\cos^2{\hat{v}})-F_{\theta}\,
(2+e\cos{\hat{v}})\sin{\hat{v}} \right].
\end{equation}
Therefore, the equations of motion in Cartesian coordinates
are equivalent to Delaunay's system given by 
equations~(\ref{Leqn}),~(\ref{Geqn}),~(\ref{elleqn}), and~(\ref{geqn}).
These equations for the planar problem are closely related to corresponding
equations for the variation of the orbital elements of the osculating
ellipse. 
In the general, i.e. three-dimensional, Kepler problem, the osculating
ellipse is characterized by six orbital elements and the
variation of these orbital elements due to perturbing forces is 
given by a set of equations that are usually
associated in celestial mechanics with the name of Lagrange 
(e.g. ``Lagrange's planetary equations'').
In general, the path of the osculating ellipse in
the six-dimensional manifold of
orbital elements encounters 
singularities at $e=0$ and $e=1$. These
singularities,
corresponding to a circle ($G=L$) and a straight
line ($G=0)$, respectively, are also evident in the
Delaunay equations.

The right hand side of the set of Delaunay equations should also be
represented in Delaunay elements.   
To accomplish this, recall equations~(\ref{Hatdefn})  and let 
$\delta f_D-\epsilon {\cal K}_D$, where $D$ denotes any one of the Delaunay
variables, be the net contribution of radiation reaction and external forces
to the rate of variation of Delaunay element $D$ with respect to time.
We have shown in a previous paper \cite{cmr} that
\begin{eqnarray}
{\cal K}_L  & = &  \frac{\partial{\cal C}}{\partial \ell} \phi(t) +
\frac{\partial{\cal S}}{\partial \ell} \psi(t), \nonumber \\
{\cal K}_{G} & = &   \frac{\partial{\cal C}}{\partial g} \phi(t) +
\frac{\partial{\cal S}}{\partial g} \psi(t), \nonumber \\
{\cal K}_{\ell} & = &- \left(\frac{\partial{\cal C}}{\partial L} \phi(t) +
\frac{\partial{\cal S}}{\partial L} \psi(t)\right), \nonumber \\
{\cal K}_g   & = & -\left(\frac{\partial{\cal C}}{\partial G} \phi(t) +
\frac{\partial{\cal S}}{\partial G} \psi(t)\right),
\end{eqnarray}
where
\begin{equation}\label{phipsieqn}
\phi(t) = \frac{1}{2}\alpha\Omega^2\cos{(\Omega t)}, \: \: \psi(t) =
\frac{1}{2}\beta\Omega^2\cos{(\Omega t+\rho)},
\end{equation}
and
\begin{eqnarray}\label{CSfs}
{\cal C}(L,G,\ell,g) & = & \frac{5}{2}a^2e^2\cos{2g} \nonumber \\
    && +a^2\sum_{\nu =1}^\infty (A_\nu\cos{2g}\cos {\nu\ell}
                            -B_\nu\sin{2g}\sin{\nu \ell}),  \nonumber \\
{\cal S}(L,G,\ell,g) & = &\frac{5}{2}a^2e^2\sin{2g} \nonumber \\
   &&  +a^2\sum_{\nu =1}^\infty (A_\nu\sin{2g}\cos {\nu\ell}
                            +B_\nu\cos{2g}\sin{\nu \ell}).
\end{eqnarray}
Here we have introduced $A_{\nu}$ and $B_{\nu}$, 
which are given in terms of the
Bessel functions by
\begin{eqnarray}\label{ABnu}
A_\nu & = &\frac{4}{\nu^2e^2}
            (2\nu e(1-e^2)J_\nu'(\nu e)-(2-e^2)J_\nu(\nu e)), \nonumber \\
B_\nu & = & -\frac{8}{\nu^2e^2}\sqrt{1-e^2}\,
     (e J_\nu'(\nu e)-\nu (1-e^2)J_\nu(\nu e)). \nonumber
\end{eqnarray}

Let us now compute the contribution of radiation damping terms to the equations
of motion.  Using equation (\ref{frftheta}) and the relationships
\[v^2 = \frac{2}{r} -\frac{1}{a},  \: \: \dot{r}^2 = -\frac{1}{a}+\frac{2}{r}
-\frac{G^2}{r^2}\:, \] we find that
\begin{eqnarray}\label{frfthetadel}
f_r & = & -\frac{4 e\sin{\hat{v}}}{GL^2r^3}
  \left(1- \frac{10}{3}\frac{L^2}{r}-5\frac{L^2G^2}{r^2}\right), \nonumber \\
f_{\theta} & = & -\frac{18 G}{L^2 r^4}\left(1-\frac{20}{9}\frac{L^2}{r}
  +\frac{5}{3}\frac{L^2G^2}{r^2}\right).
\end{eqnarray}
The substitution of equation~(\ref{frfthetadel}) into 
the Delaunay equations via equation~(\ref{Hatdefn})
results in the set of equations presented in equation~(\ref{feq1}),
where $e = \sqrt{1-G^2/L^2}$;
moreover, these relations 
involve powers of $1/r$ as well as
$\sin{\hat{v}}$ times powers of $1/r$.  Therefore, to complete the discussion
we need Fourier expansions of these expressions in terms of $\cos{\nu\ell}$ and
$\sin{\nu\ell}$ similar to those given in equation (\ref{CSfs}).  
This can be accomplished by extending the classical methods of celestial
mechanics described, for instance, in the definitive
monograph of Watson~\cite{watson}. The final expressions for
$f_D$, $D\in\{L,G,\ell,g\}$, are not used explicitly in this paper; therefore,
we do not present them here for the sake of brevity.

Finally, let us note that the Delaunay equations 
can now be expressed in the form
given in equation~(\ref{D2EQM1}).
\section{Average Rate of Damping}\label{RadForceApp}
The average contribution of radiation damping to the equations
of motion in Delaunay variables is employed
in this paper in the calculation of the bifurcation function
(Section~\ref{conpo}) as well as in the description of averaged dynamics
(Section~\ref{avedyn}). In the light of the results of the previous
appendix, the quantity $f_D$, where $D$ is any one of the Delaunay
variables, is given by equation~(\ref{D2EQM1}) and can be expressed
as a Fourier series
\begin{equation}\label{AveBasicEq}
f_{D}(L, G, \ell, g) = 
   a_0 +\sum_{\nu = 1}^{\infty}(a_{\nu}\cos{\nu\ell}+b_{\nu}\sin{\nu\ell}).
\end{equation}
We are interested in $\bar f_D$, which is the average of $f_D$ with
respect to the ``fast'' angular variable $\ell$, i.e.   
\[
\bar f_D:=\frac{1}{2\pi}\int_0^{2\pi}f_D(L,G,\ell,g)\,d\ell.
\]
It is clear that $\bar f_D=a_0$, and this is the quantity that is needed 
in Section~\ref{avedyn}.

The contribution of radiation damping to
the bifurcation function 
$\cal B$ is given by $\Delta{\cal F}_{D}$, where
\begin{equation}\label{BifAveFun}
{\cal F}_{D} = \int_0^{2\pi m/ \Omega} f_{D}(L, G, \omega t + \ell, g)
\: dt. 
\end{equation}
Using the change of variable $\tilde t = \Omega t/m + \ell/n$, the resonance
condition $\omega = n\Omega /m$, and the periodicity of $f_{D}$ 
with respect to $\ell$, equation (\ref{BifAveFun}) can be written as
${\cal F}_{D}=2\pi n L^3 \bar{f}_D$. 
It remains to compute $\bar{f}_D$.

Using the relation $d\ell = r^2 d\hat{v}/(a^2\sqrt{1-e^2})$, 
we can write
\begin{equation}\label{87}
\bar{f}_{\cal D} = \frac{1}{2\pi a^2\sqrt{1-e^2}}
                 \int_0^{2\pi} r^2{\cal F}_{D}\,d\hat{v}=
          \frac{1}{L^3 G}<r^2 f_D>,
\end{equation}
where the angular brackets denote the average over the true anomaly $\hat{v}$. 
The various averages may be evaluated using
\[
G^2 \left< \frac{1}{r} \right>  =  1,\quad 
G^4 \left< \frac{1}{r^2} \right>  =  1+\frac{1}{2} e^2,\quad 
G^6 \left< \frac{1}{r^3} \right>  =  1+\frac{3}{2} e^2,  
\]
\[
G^8 \left< \frac{1}{r^4} \right>  =  1+3 e^2 + \frac{3}{8} e^4, \quad
G^{10} \left< \frac{1}{r^5} \right>  =  1+ 5 e^2 + \frac{15}{8} e^4, 
\]
\[
G^2 \left< \frac{\sin^2{\hat{v}}}{r} \right>  =  \frac{1}{2},  \quad 
G^4 \left< \frac{\sin^2{\hat{v}}}{r^2} \right>  =   \frac{1}{2}\left(
1+\frac{e^2}{4} \right),   
\]
\begin{equation}\label{AveValues}
G^6 \left< \frac{\sin^2{\hat{v}}}{r^3} \right>  =  
 \frac{1}{2}\left( 1+\frac{3}{4}e^2 \right).
\end{equation}
We find that
\begin{eqnarray}\label{fqAve3}
\bar f_{L}&=&-\frac{1}{G^7}\big( 8+\frac{73}{3} e^2+ \frac{37}{12} e^4 \big),
\nonumber \\
\bar f_{G}&=&-\frac{1}{L^3G^4}(8+7e^2),\nonumber \\
\bar f_{\ell}&=&0, \qquad \bar f_{g}=0,
\end{eqnarray}
where the last two equations simply follow from the fact that
\[<h(\cos{\hat{v}})\,\sin{\hat{v}}>=0\] for every continuous function $h$.

For the bifurcation function $\cal B$, only ${\cal F}_L$, ${\cal F}_G$, and 
${\cal F}_g$
are needed and these are then given by
\begin{eqnarray}\label{fqAve2}
{\cal F}_L& =& 
  - 2\pi n\frac{L^3}{G^7}\left( 8+\frac{73}{3} e^2+ \frac{37}{12} e^4 \right),
\nonumber \\
{\cal F}_G &=& -2\pi n\frac{1}{G^4} (8+7e^2), \nonumber \\
{\cal F}_g & = & 0.
\end{eqnarray}
%\begin{thebibliography}{99}

%\newpage
%\centerline{Figure Captions}
%\vspace{2in}
\begin{figure}[h]
\vspace*{1in}
\caption[]{
The plots are for system~(\ref{NumExpEq}) with parameter values
$\epsilon=10^{-4}$, $\delta/\epsilon=10^{-3}$, 
$\alpha=2$, $\beta=2$, $\rho=0$, and $\Omega= .6495190528$.
The top panel shows $L=\sqrt{a}$ versus time for the initial conditions 
$(p_r,p_\theta,r,\theta)$
equal to $(.38927, .61085 , 2.70900, -2.45677)$.  The middle
panel shows $L$ versus time and the bottom panel shows $G$ versus $L$
for the initial conditions $(.3893, .6109 , 2.709, -2.4568)$. 
Here, $L\approx 1.1547$ corresponds to $(1:1)$ resonance,
$1/L^3=\Omega$.
\label{capfig}}
\end{figure}
\begin{figure}[h]
\caption[]{
The plot of $G$ versus $L$ for system~(\ref{NumExpEq}) with
initial conditions $(p_r,p_\theta,r,\theta)$
equal to $(.5, 1 , 1, 0)$ and parameter values
$\epsilon=10^{-3}$, $\delta/\epsilon=10^{-5}$,
$\alpha=2$, $\beta=2$, $\rho=0$, and $\Omega=1.299038106$.
The top panel depicts 157000 points, one after each time interval
$2 \pi/\Omega$.
The middle panel is a blow up for iterates 110000--130000, the bottom
panel for 140000--155000.
\label{trans}}
\end{figure}
\end{document}